\documentclass[conference]{IEEEtran}
\pdfoutput=1
\IEEEoverridecommandlockouts
\usepackage{cite}
\usepackage{amsmath,amssymb,amsfonts}
\usepackage{algorithmic}
\usepackage{graphicx}
\usepackage{textcomp}
\usepackage{xcolor}
\def\BibTeX{{\rm B\kern-.05em{\sc i\kern-.025em b}\kern-.08em
    T\kern-.1667em\lower.7ex\hbox{E}\kern-.125emX}}
    
\usepackage{multirow}
\usepackage{hhline}
\usepackage{multicol}
\usepackage{enumitem}
\usepackage{lipsum}
\usepackage{url}

\AtBeginDocument{%
  \providecommand\BibTeX{{%
    \normalfont B\kern-0.5em{\scshape i\kern-0.25em b}\kern-0.8em\TeX}}}
    
    \newenvironment{boxedtext}
    {
    
    \begin{center}

    \begin{tabular}{|p{0.96\linewidth}|}
    \hline
    }
    { 
    \\ \hline
    \end{tabular} 
    
    \end{center}
    \vspace{5pt}
       }
\begin{document}

\title{A Benchmark Study of the Contemporary Toxicity Detectors on Software Engineering Interactions}

\author{\IEEEauthorblockN{Jaydeb Sarker, Asif Kamal Turzo, Amiangshu Bosu}
\IEEEauthorblockA{\textit{Department of Computer Science} \\
\textit{Wayne State University}\\
Detroit, MI, USA \\
\{jaydebsarkar, asifkamal, amiangshu.bosu\}@wayne.edu}

}

\maketitle

\begin{abstract}
%Toxic conversations during software development interactions may have serious repercussions on an Open Source Software (OSS) development project. For example, a victim of toxic conversation may become afraid to express him/herself, therefore get demotivated, and may  eventually leave the project. 
Automated filtering of toxic conversations may help an Open-source software (OSS)   community to maintain healthy  interactions among the project participants. Although, several general purpose tools exist to identify toxic contents, those may incorrectly flag some words commonly used in the Software Engineering (SE) context as toxic (e.g., `junk', `kill', and `dump') and vice versa. To encounter this challenge, an SE specific tool has been proposed by the CMU Strudel Lab (referred as the `STRUDEL' hereinafter) by combining the output  of the Perspective API with the output from a customized version of the Stanford's Politeness detector tool. However, since STRUDEL's evaluation was very limited with only 654 SE text, its practical applicability is unclear. Therefore, this study aims to empirically evaluate the Strudel tool as well as four state-of-the-art general purpose toxicity detectors on a large scale SE dataset. On this goal, we empirically developed a rubric to manually label toxic SE interactions. Using this rubric, we manually labeled a dataset of 6,533 code review comments and 4,140 Gitter messages. The results of our analyses suggest significant degradation of all tools' performances on our datasets. Those degradations were significantly higher on our dataset of formal SE communication such as code review than on our dataset of informal communication such as Gitter messages. Two of the models from our study showed significant performance improvements during 10-fold cross validations after we retrained those on our SE datasets.  Based on our manual investigations of the incorrectly classified text, we have identified several recommendations for developing an SE specific toxicity detector.
\end{abstract}


\section{Introduction}
\label{sec:intro}

%\anote{Context: Describe why identification of toxic text from software developer interactions is important. Don't forget to add few citations. One or two paragraphs. }
Prior research have found multiple evidence of toxic interactions, such as: profanity, insult, hate speech, identity attack, misogynistic remarks, flirtations, or sexual innuendos, among several Free / Open Source Software (FOSS) projects~\cite{squire2015floss,toxic-blog,nafus2006gender,nasif-icse2019,paul-SANER-2019}. Toxic  interactions  may have serious repercussions on a FOSS project. For example, a victim of toxic conversation may become afraid to express him/herself, therefore get demotivated, and may  eventually leave the project. 

The problem of online toxic conversations are more widespread than the FOSS projects. For example, a 2017 survey conducted by the Pew Research Center found that two out of five Americans have experienced online harassment~\cite{duggan2017online}. Another study found 43\% college students reporting being recipients of harassing messages~\cite{lindsay2016experiences}. More than one-third victims of  those abusive online interactions reported feeling depressed~\cite{lindsay2016experiences}. Therefore, researchers have been focusing on automatic identification  of toxic online conversations~\cite{perspective-api,kurita2019towards,gunasekara2018review,d2020bert,alshemali2020improving} to prevent such negative incidents. Jigsaw (a unit of Google)\footnote{\url{https://jigsaw.google.com}} is on the forefront of this research  with building a public API named the Perspective API\footnote{\url{https://www.perspectiveapi.com}} to automatically score perceived toxicity of a text. Kaggle\footnote{\url{https://www.kaggle.com/}}, in collaboration with the Jigsaw, started a competition called `Toxic Comment Classification Challenge in 2018' \footnote{\url{https://www.kaggle.com/c/jigsaw-toxic-comment-classification-challenge}} with the goal of building a classifier that can classify toxic contents better than the Perspective API. This challenge have produced  high quality toxicity detectors with AUC \footnote{Area under receiver operating characteristic curve represents ability of of a binary classifier system as its discrimination threshold is varied.} scores (i.e.,  0.988).

Since software development is a collaborative activity,  toxic conversations within a team may not only degrade relationships among team members but also have a great impact on the productivity of a developer~\cite{bosu2013impact}. 
Fear of bullying can refrain a developer from sharing his/her opinions or discourage a newcomer from seeking expert suggestions.
Prior studies have found developers expressing frustrations over peers with `prickly' personalities~\cite{Bosu-EMSE-2019-Blockchain,filippova2016effects}.  Toxic conversations may not only demotivate developers but also waste valuable work hours~\cite{ramanstress}.  Since software development communities, such as the FOSS projects, are professional communities, automated identifications of toxic conversations from software developer communications are crucial.
%Toxic code review comments not only make a bad relationship between persons but also reduce the productivity of developers. 

 However, as prior research on building sentiment analysis tools for the Software Engineering (SE) domain has shown~\cite{jongeling2017negative}, toxicity detectors developed for other domains may not work well  on SE conversations. An off-the-shelf toxicity detector may incorrectly flag some words commonly used in the SE context as toxic (e.g., `junk', `kill', and `dump'). To encounter this challenge, an SE specific tool has been proposed by the CMU Strudel Lab (referred as the `STRUDEL tool' hereinafter) by combining the output  of the Perspective API~\cite{perspective-api} with the output from a customized version of the Stanford's Politeness detector tool~\cite{danescu2013computational}. Since the STRUDEL had a low F-score (0.57) during its limited evaluation with only 654 SE text, its practical applicability is uncertain.  
Therefore, this study aims \textit{to empirically evaluate the STRUDEL tool as well as four other state-of-the-art general purpose toxicity detectors on a large scale SE dataset. }
On this goal, we  empirically developed a rubric to manually label each SE text as either `toxic' or `non-toxic' and using that rubric manually labeled two datasets of: i) 6,533 code review comments and ii) 4,140 Gitter messages. Using these datasets, we evaluated the five selected tools and answer the following three research questions:

\begin{enumerate}[start=1,label={(\bfseries RQ\arabic*):}]

\item \textbf{How do contemporary toxicity detectors perform on an SE dataset?}

\underline{Motivation:} This evaluation will enable us to identify the practical applicability of contemporary toxicity detectors on SE conversations.

\underline{Results:} None of the contemporary toxicity detectors achieved adequate performances to justify practical applications. The F-scores of the tools included in our study dropped more significantly on a formal SE communication dataset such as code review than on a informal communication such as Gitter messages. The significant disagreements between most of the tool pairs also suggest that results of an empirical study using one of these tools may differ significantly if we switch from one tool to another one.

\item  \textbf{What are the categories of SE texts that contemporary toxicity detectors are more likely to misclassify?}

\underline{Motivation:} This analysis will identify scenarios to consider when developing a customized toxicity detector for the SE domain.

\underline{Results:} Most of the tools accurately identifies texts with expletives or swear words. However, those tools fails on words that have different meaning in the SE context. Moreover, contemporary tools also fail on sentences expressing humility, where an author uses demeaning words (e.g., `stupid', `dumb', and `idiot') referring him/herself or his/her own work.

\item \textbf{ Does retraining on a SE dataset improve the performances of contemporary toxicity detectors?}

\underline{Motivation:} This analysis will show us the easeness or difficulty in building a customized toxicity detector for the SE domain.

\underline{Results:} The results are highly promising with both of our retrained models outperforming contemporary models during 10-fold cross- validations on our SE datasets. A large scale labeled toxicity dataset of SE interactions may enable the development of reliable SE domain specific toxicity detectors.

\end{enumerate}

\textbf{The primary contributions} of this paper are:
\begin{itemize}
 \item An empirically developed rubric to manually label the toxicity of SE conversations.
 
 \item Two manually labeled toxicity datasets from the SE domain, with one dataset including 6,533 code review comments and the other dataset including 4,140 Gitter messages.

\item An empirical evaluation of five contemporary toxicity detectors on two SE datasets.

\item Empirical evidence depicting the possible development of a reliable toxicity detector for the SE domain by retraining exiting models on a SE dataset.

\item  A set of guidance for SE researchers on building customized toxicity detectors for the SE domain.

\item \textbf{Enabling replication:} To enable future studies, we have made our dataset and results available online at: \url{https://github.com/WSU-SEAL/toxicity-dataset} 
 %\item There is no labeled dataset for toxic content in software engineering. We have manually labeled code review (4,212) and Gitter messages (2,949). 
%\item We have implemented six state of art tools to generate the toxic score of our labeled dataset. Then we studied the performance of six tools.
    
%\item We also studied the about the toxic interactions among developers both code review and Gitter communications. This would be helpful for further direction in emotional interactions among software communities. 

%\item We discuss the challenges to build an automatic toxic content classifier in SE domain. 
\end{itemize}

%\anote{Outline: An outline of the paper}
\textbf{Paper organization:} The remainder of this paper is organized as following.
Section~\ref{background} provides a brief background on prior research in identifying online toxic contents. 
Section~\ref{method} details our research methodology.
Section~\ref{sec:evaluation} describes the results of our empirical evaluation.
%discusses the research direction and evaluates the existing tools for toxic content classification. 
Section~\ref{discussion} discusses the considerations for building an SE domain specific toxicity detector. 
Section \ref{threats} discusses the threats to validity of our findings. 
Finally, Section \ref{conclusion} provides the future direction of our work and concludes this paper. 
\section{Background}
\label{background}
Following subsections provide a definition of toxic contents in the context of the SE domain and  briefly describe prior research in identifying online toxic contents. 
% This section needs extensive citations. Possibly each sentence should be justified by a citation.
%Toxic content classification is a task of sentence classification in natural language processing. There are some works done in toxic content classifications. In our study, we find that traditional classifiers do not perform well in code review comments. Code review comments consist of technical terms and code snippets. Traditional classifiers miss classify the comments. One potential reason behind that is there are several jargon in software engineering domain that can be portrayed as negativity in real world discussion (i.e., “kill” the process).

%\anote{What is toxicity analysis?}
%\jnote{definition and related work}
\subsection{Toxic Contents}
\textit{Toxicity analysis} is a natural language classification problem of identifying toxic contents. However, prior research had differing views on  which contents should be considered as toxic. For example, the pew research center  classifies offensive name calling, threats,  and sexual harassment~\cite{duggan2017online} as toxic interactions. Zaheri \textit{et} al.  includes insult, verbal sexual harassment, threats, obscene languages  in their analyses of toxicity~\cite{zaheri2020toxic}. Georgakopoulos \textit{et} al. included personal attacks, online harassment and bullying behaviors among toxic interactions \cite{georgakopoulos2018convolutional}. Kurita \textit{et} al. expanded their definition of  toxic contents by including texts that can be harmful or offending to the recipient(s) \cite{kurita2019towards}. The Perspective API~\cite{perspective-api} defines toxic contents  as ``texts that are rude, disrespectful, or unreasonable''. Since the SE domain consists of professional communities, we adopt the following expansive definition of toxic contents in our analyses:

\begin{quote}
 \textit{{``An SE conversation will be considered as toxic, if it includes any of the following: i) offensive name calling, ii) insults, iii) threats, iv) personal attacks, v) flirtations, vi) reference to sexual activities, and vii) swearing or cursing.''}}
\end{quote}

%They also identified top 100 toxic lexicons in English language. 

%\anote{Brief overview of the approaches taken by exiting tools. }
%\jnote{existing tools overview}

\subsection{Prior Research on Toxicity Analysis}

The Jigsaw team from Google developed the Perspective API to identify abusive online contents, which is considered one of the contemporary state-of-the-art tools. The models behind the Perspective API (PPA) are trained using manual annotations from crowd sourced human raters based on a published guidelines~\cite{jigsaw-guideline}.
%listed at: \url{https://github.com/conversationai/conversationai.github.io/blob/master/crowdsourcing_annotation_schemes/toxicity_with_subattributes.md}. 
Besides the toxicity model, PPA also provides experimental models  to identify insults, profanities, identity\_attacks, sexually\_explicit contents, flirtations, and threats in a text. However, PPA is not free from flaws, as  Hosseini \textit{et} al.  pointed out tricks to deceive the PPA~\cite{hosseini2017deceiving}.

Georgakopoulos et al. proposed a Convolutional Neural Network (CNN)  based model trained on the Jigsaw dataset \cite{georgakopoulos2018convolutional} and found that CNN based models performed better than bag-of-words based models in identifying toxic texts. Recently, researches have proposed several deep learning based toxicity classifiers~\cite{chen2019use,elnaggar2018stop,srivastava2018identifying} and provided guidelines on improving performances of these models.

Identifications of obfuscated or perturbed toxic words (e.g., `fuc\_k, and `idoit') have been a major limitation of toxicity classifiers. Mishra et al. proposed a single embedding for unseen words to classify the toxic context, which is  able to identify obfuscated and non-obfuscated words \cite{mishra2018neural}. 
Kurita et al. proposed a model called Contextual Denoising Autoencoder (CDAE) to classify toxic contents \cite{kurita2019towards}. They leveraged both  character-level and contextual information  in their models to overcome the limitations of contemporary toxicity classifiers in identifying perturbed toxic tokens (e.g., intentional typos to evade detection). 

Due to the biases among the human raters against certain group of people (e.g., gay, lesbian, black, and muslim), toxicity classifiers based on several dataset are biased against those groups~\cite{vaidya2019empirical}. To overcome this challenge, Vaidya et al. proposed a multi-task learning model to reduced identity  biases among toxicity classifiers~\cite{vaidya2019empirical}. 

Identifications of online bullying and hate speeches have also been focuses of several prior works. Waseem and Hovy proposed a set of 11 rules to annotate hateful tweets and labeled more than 16K public tweets using these criteria~\cite{waseem2016hateful}. Using this dataset,  they prepared a dictionary  of the most hateful indicative words.  Davidson et al. proposed a classifier to identify both hate speeches and offensive languages~\cite{davidson2017automated}.
Chandrasekharan et al. proposed a model named Bag of Communities to identify abusive online interactions, where they showed that models trained based on  labeled data obtained from one online community can be successfully reused to identify abusive contents from a different community~\cite{chandrasekharan2017bag}.
%Other prior works in  There are some several works done to detect hate speech form online communications \cite{kwok2013locate} \cite{burnap2015cyber}. Bag of Communities (BoC) is a model to detect the abusive behavior from online communication where preexisting data from other communities is not needed to train the models \cite{chandrasekharan2017bag}.

%\anote{Is there any work in the SE domain on this topic}

Although, toxic contents have been found among SE interactions~\cite{squire2015floss,toxic-blog,nafus2006gender,nasif-icse2019,paul-SANER-2019}, most of the prior works in the SE domain focused on identifying software developers' sentiments\cite{ahmed2017senticr,calefato2018sentiment,islam2018sentistrength}. 
Raman et al. proposed the first SE domain specific toxicity detector, and used that to study unhealthy interactions among FOSS developers~\cite{ramanstress}. 
%They proposed a SE specific tool (Strudel tool) to identify toxic comments in SE domain. Their work focus on the recent years communications of GITHUB communities. There are not such other works in toxicity analysis in Software Engineering domain. Researhcers in SE domain proposed some customized tools to detect the sentiment in SE dataset \cite{ahmed2017senticr} \cite{calefato2018sentiment} \cite{islam2018sentistrength}. 
\section{Research Method} \label{method}
Currently, the STRUDEL dataset of 654 texts~\cite{ramanstress} is the only labeled toxicity dataset from the SE domain. Due to the small size of the STRUDEL dataset, we decided to create a new labeled toxicity dataset from SE interactions.
Following subsections describe our approach to select text for our dataset, our manual labelling process, tool selection, and empirical evaluation of the selected tools.

\subsection{Data Source Selection}
Since toxic interactions are not very frequent among FOSS projects~\cite{carolyn-icse-2020,ramanstress}, we looked into prior research and identified following two mediums, where toxic conversations are more likely to occur. 
 \begin{itemize}
     \item \textbf{Code review} is a software development practice, where a developer sends his/her changes to peers for manual reviews. Although, code review is a formal process,  incidents of profanity or insults are not uncommon during code reviews.~\cite{paul-SANER-2019,carolyn-icse-2020}. We selected three popular FOSS projects (i.e., Android, Chromium OS, and LibreOffice) as our data sources, since a recent research suggests toxicity among code reviews of those projects~\cite{paul-SANER-2019}.
 
     \item \textbf{Gitter} is an open-source instant messaging and chat room system  for software developers. Although Gitter is similar to Instant Relay Chat (IRC), Gitter's integration with Github repositories has made it popular among recent FOSS projects. Since prior research found toxic interactions among FOSS IRC channels~\cite{squire2015floss,crowston2017lessons}, we considered chat messages as one of our sources. We selected the gitter channel of the Ethereum project~\footnote{\url{https://gitter.im/ethereum/go-ethereum}}, since it is one of the most active channels on Gitter.
 \end{itemize}

\subsection{Data Mining}
The code review repositories of the  three selected projects are managed by Gerrit\footnote{https://www.gerritcodereview.com/}. We wrote a Python script to access Gerrit's REST API to mine all the publicly available code reviews for the three projects and store the data in a MySQL database. Using an approach similar to Paul et al.~\cite{paul-SANER-2019}, we  identified the bot accounts to exclude the comments not written by humans.
We used the GitterPy\footnote{\url{https://github.com/myuz/GitterPy}} library to connect to Gitter's REST API and download all the messages to our MySQL database. Table~\ref{tab:projects} shows an overview of the messages mined by our scripts from the four FOSS projects.

\begin{table*}
    \caption{An overview of the Messages mined by our mining scripts}
    \label{tab:projects}
    \centering
    \resizebox{\linewidth}{!}{% <------ Don't forget this %
\begin{tabular}{|l|l|l|r|r|}
\hline
        \textbf{Project} & \textbf{Data Source} & \textbf{Time period} & \textbf{Total messages} & \textbf{Toxic messages*}  \\ \hline
        Android &  Code review: \url{https://android-review.googlesource.com/} & December 2008 to June 2019 & 152,065 & 647 \\ \hline
        Chromium OS &  Code review: \url{https://chromium-review.googlesource.com/} & April 2011 to  March 2020 &   1,176,642& 2,485 \\ \hline
        LibreOffice  &  Code review: \url{https://gerrit.libreoffice.org/} & March 2012 to June 2019 & 12,273 & 81 \\ \hline
        Ethereum &  Gitter: \url{https://gitter.im/ethereum/go-ethereum} & June 2014 to March 2020 &  122,355 & 1,950 \\ \hline
        \multicolumn{5}{l}{*As classified by the Perspective API}
    \end{tabular}
    }
\end{table*}

\subsection{Dataset Generation}
Due to  the rarity of toxic interactions, a fully-randomized selection of text from our data sources would create a highly unbalanced dataset of less than 1\% toxic texts. To overcome this challenge, we adopted a  customized stratified sampling strategy~\cite{sarndal2003model} by leveraging the Google's Perspective API. First, we use the PPA to compute the toxicity score of each text. The PPA score for a text varies between 0 to 1, which indicates the probability of that text being toxic. Since  a PPA score of 0.5 or above suggests a text as more likely to be `toxic' than to be `non-toxic', we included all the texts with PPA scores above 0.5. Based on this selection, we obtained 3,213 code review texts and 1,950 Gitter messages.

Second, we divided the texts  with PPA scores  less than 0.5  from our two datasets into five equally spaced PPA score groups with each group spanning an interval of 0.1. From the code review dataset, we randomly selected 664 text from each group. For example, we randomly selected 664 code review comments with a PPA score between 0 to 0.1, 664 texts with a PPA score between 0.1 to 0.2 and so on.  Using this stratified sampling, we selected additional 3,320 code review comments that are classified as `non-toxic' by PPA. Similarly, we selected additional 2,190 gitter messages (i.e., 438 messages from each group) with PPA scores less than 0.5. 
%Our goal of this sampling was to reduce tool bias This selection stratified sampling ensured  and 1,000 Gitter messages. 

In addition to these two datasets, we randomly sampled 2,000 `non-toxic' and `1,000' toxic texts from the labeled Jigsaw test dataset~\cite{Thain2017}. We use this dataset of 3,000 texts to evaluate baseline performances of the selected tools on a non-SE dataset.

\subsection{Manual Labeling}
During the first stage of our manual labeling, we focused on developing a rubric to manually label the toxicity class of the selected texts. Our initial rubric was based on the guidelines published by the Conversation AI team~\cite{jigsaw-guideline}.  Two of the authors independently went through 1,000 texts to prepare a set of rules. Then, we had a discussion session to create a unified set of rules for labeling. Using this set of rules, two of the authors independently labeled all the selected texts. 
%Our rubric building process was continuous, as we added more rules after discussion, if we found multiple similar cases during our manual labeling that fitted none of our existing rules. 
Table~\ref{tab:rubrics} shows the set of  rules with examples taken from our dataset.

\begin{table*}
    \caption{Set of rules for classifying a text as either `toxic' or `non-toxic' with examples taken from our dataset}
    \label{tab:rubrics}
    \centering
    \resizebox{\textwidth}{!}{    
    \begin{tabular}{|l|p{4.5cm}|p{4.5cm}|p{4.5cm}|} \hline
     \textbf{\#} &\textbf{Rule}     & \textbf{Rationale} & \textbf{Example*} \\ \hline
    \textit{Rule 1:} &Inclusion of a profane or curse words in a sentence would be marked as `toxic'.
     & Profanities are the most common sources of online toxicities.   & ``we don't want to fuck 64-bit bit up like 32-bit was fucked.'' \\  \hline
  
    \textit{Rule 2:} & Inclusion of an acronym, that refers to expletive or swearing, in a sentence would be marked as `toxic'.   & Sometimes people use acronyms of profanities, which are equally toxic as its' expanded form.  & ``wtf is going on with this nonstop?'' \\ \hline
   \textit{ Rule 3:} &Insult to another person or person's work would be marked as `toxic'. & Insulting another developer may create a toxic environment and should not be encouraged.  & ``YOU MUST BE A BIG FOOL''\\ \hline
    
    \textit{Rule 4:} &Attacking  a person's identity  (e.g., race, religion, nationality, gender or sexual orientation) would be marked as `toxic'. &  Identity attacks are considered toxic among all categories of online conversations. &  ``you are twice as smart as a typical stupid American consumer, you get to have an unlimited number of children''\\ \hline
    
     \textit{ Rule 5:}&  Threatening another person or a community would be marked as `toxic'.  & Threats may stir hostility between two developers and force the the recipients leave the community.   &  ``One of these days I’m going to slap you @****'' \\ \hline
    
    \textit{Rule 6:} &Both implicit or explicit References to  sexual activities would be marked as `toxic'. &  Implict or explicit references to sexual activities may make some developers, particularly females, uncomfortable and make them leave a conversation. & ``i know but...well its like masturbating vs sex you see what i mean
'' \\ \hline
    
     \textit{Rule 7:} &Flirtations would be marked as `toxic'. & Flirtations may also make a developer uncomfortable and make a recipient avoid the other person during future collaborations  & ``Just like how I told that woman I thought she looked pretty.
'' \\  \hline
    
   \textit{ Rule 8:} & If a demeaning word (e.g., `dumb', `stupid', `idiot', `ignorant') refers to either the writer him/herself or his/her work, the sentence would not be marked as `toxic', if it does not fit any of the first seven rules.  & It is common in SE community to use those word for expressing their own mistakes. In those cases, the use of those toxic words to himself/herself does not make toxic meaning.    & ``stupid me, my editor shows them the same color and tricks me every time.''  \\ \hline
   
%     \textit{ Rule 9:} & Sentences with the words, such as `nerd', and `geek' would not be marked as `toxic', if a sentence does not violate any of the first seven rules. & Though these words have negative origins, tagging a person with these words  are no long considered toxic in the SE domain.  & ``nerd snipe of trailing white space.''  \\ \hline

\textit{ Rule 9:} & A sentence, that does not fit rules 1 through 8, would be marked as `non-toxic'. & General non-toxic comments.   & ``you can delete this main logic as the wrapper.py handles it for you.'' \\  \hline

\multicolumn{4}{p{15cm}}{* \textit{Examples are provided verbatim, to accurately represent the context. We did not censor any text, except omitting the reference to a  person's name.}}\\

    \end{tabular}
}
\end{table*}

After the independent manual labeling, we compared the labels from the two raters to identify conflicts. Out of the 6,533 comments from the code review dataset, the two raters agreed on 5,950 comments (i.e., 91.1\%). On the Gitter dataset, out of the 4,140 messages, the raters agreed on 3,730 messages (i.e., 90.1\%). We also measured the level of agreement between the two raters using Cohen's Kappa ( $\kappa$) \cite{cohen1960coefficient}, which was estimated as 0.727 for our code review dataset and as 0.781 for our Gitter dataset. Kappa ($\kappa$) values  are commonly interpreted as follows: values $\leq$ 0 as indicating `no agreement' and 0.01–0.20 as `none to slight', 0.21–0.40 as `fair', 0.41– 0.60 as `moderate', 0.61–0.80 as `substantial', and 0.81–1.00 as `almost perfect agreement'. Therefore, the level of agreement between the two raters during our manual labeling can be considered as `substantial'.

%to measure the agreement between two raters. It shows  p-value = 0, z = 47.4 and  Kappa score is 0.731 for unweighted cohen kappa. It seems a good agreement between two raters.
%Similarly, we found z = 40.4, p-value = 0 and Kappa score  0.737 for code review dataset with two raters. The agreement of between two raters after indepentdently labeling is  86.9\% (2563) comments. The meaning of toxicity sometimes varies from person to person. For that reason, we found total 847 (461 for code review, 386 for gitter ethereum dataset) comments where two raters do not agree with each other. 
Finally, we had discussion sessions to review the disagreements and come up  with an agreed upon rating for each of the texts with a conflicted rating. After the conflict resolution process, we found 20\% `toxic' texts in our code review dataset, while the Gitter dataset had 35.4\%. Since real-time chats are bit more informal than code reviews, a higher ratio of toxic texts in the Gitter dataset may not be surprising. Table~\ref{tab:dataset-overview} provides a brief overview of the Jigsaw sample dataset as well as the  two SE datasets after the completion of our manual labeling.

%After completing the labeling, we have found that 1274 (30\%) comments are toxic and 2938 (70\%) are non-toxic comments in code review dataset. On the other hand, gitter ethereum has 1379 (47.3\%)  toxic comments and 1552 non-toxic comments. Table \ref{tab:tab1} shows the values of final labeling. We also observe that developers use more toxic comments than the code review comments. Gitter is a online platform where software developers communicate with each other. It is common for them to use toxic language to communicate with each others. On the other hand, code reviewers rarely use toxic comments in their comments.   

\begin{table}[]
\caption {An overview of the three datasets} \label{tab:dataset-overview} 
\centering
\begin{tabular}{|l|r|r|r|}
\hline
 \textbf{Dataset} & \textbf{\# total texts} &  \textbf{\# toxic} & \textbf{\# non-toxic} \\ \hline
 Jigsaw Sample & 3,000 & 1,000 & 2,000 \\ \hline
  Code Review & 6,533 & 1,310 & 5,223 \\ \hline
 Gitter Ethereum& 4,140 & 1,468 & 2,672        \\ \hline
 
\end{tabular}

\end{table}

\subsection{Tool Selection}
%\anote{ Provide a brief overview of each tool. It should briefly describe how the tool was trained and which approach it uses. What was the performance reported in the paper (if reported).} 
We selected total five tools for evaluation. We selected the Perspective API~\cite{perspective-api}, since it has been widely used, and is considered one of state-of-the-art tool for toxic text classification. The STRUDEL tool~\cite{ramanstress} was selected, since it is the only SE domain specific toxicity detector. The remaining three tools for evaluation were selected based on following two criteria. 

\begin{enumerate}
    \item The design and evaluation of the tool was published as a research paper.
    \item The source code of the tool is publicly available for download and evaluation.
\end{enumerate}

 Although, we noticed an influx of toxicity detector publicly available on the Github (\url{https://github.com/topics/toxic-comment-classification}) as a part of the 2018 Kaggle classification challenge~\cite{kaggle2018}, most of those tools fail our first inclusion criteria. In the following subsections, we provide a brief overview of the five tools selected for our analyses.

\subsubsection{Perspective API (PPA)}

The developers from the Google's Conversational AI and Jigsaw  developed the Perspective API to identify abusive online contents~\cite{perspective-api}. The primary goal of the PPA is to develop a online platform to reduce the toxic comments and create spaces for healthy conversations\footnote{\url{https://jigsaw.google.com/}}.
The Perspective API currently provides six different models to rate the level of toxicity, profanity, insult, identity attack, threat, sexual explicit from a text. 
 %Jigsaw team is also working to identify and reduce the violent extremism in online platform. The developers trained the Perspective API by asking people to label the comments as very toxic to very healthy. A comment that consists rude, disrespectful, or unreasonable is considered as toxic comment. Though the API gives the score of 8 categories of attack, we consider only the score of toxicity. 
 The PPA scores for a text from the six models vary from 0 to 1, which indicate the the probability of that text belonging to  a particular category (e.g., toxic, threat, or insult)~\cite{wulczyn2017ex}. We use the publicly available REST API to compute PPA scores for each text in our datasets. Using the score from the PPA `toxicity model', we classify a text as toxic if it has a PPA score of 0.5 or higher.

\subsubsection{STRUDEL Toxicity Detector (STRUDEL)}
Raman et al. proposed the first toxicity detector for the  SE domain recently \cite{ramanstress}. They observed that many texts were incorrectly classified as `toxic' by the PPA due to the occurrences of words that are considered `toxic' in non-SE context but are included in the technical SE vocabulary (e.g., kill, abort, and die). They developed an automated pre-processing model to identify  words that are significantly over-represented in a SE dataset compared to general English and replace those words with more neutral filler words. They used a modified version of the Stanford's politeness detector tool~\cite{danescu2013computational} to identify the politeness score of the pre-processed text. Finally, they used an SVM classifier to classify the text as either `toxic' or `non-toxic' by combining the PPA score of the unmodified text with the politeness score of the pre-processed text.    

The STRUDEL tool was evaluated using a dataset of 654 issue comments mined from Github, where only 167 comments were labeled as `toxic'. During their evaluation it achieved  a precision of 0.91 and a recall of 0.42 of their test dataset. Moreover, on a  dataset of 100,000 randomly sampled GitHub issues it achieved 50\% precision in identifying `toxic' comments. In our evaluation, we use the pretrained STRUDEL model available on Github.

%They developed an SVM classifier to detect the unhealthy interactions in software communities. First, they manually labled  386 issue threads of GitHub as toxic or non-toxic. They make their classifier by combination of pretrained Perspective API, stanford politeness tool, sentiment detector for social media \cite{hutto2014vader} and abusive language detectors. Finally, they trained the SVM classifier to classify a comment as toxic or not. Moreover, they also tune the the comments which contains technical phrases. For example `kill' and `abort' are toxic terms in public domain but neutral term in software domain. They replaced those technical words by neutral words using log odds with Dirichlet prior \cite{monroe2008fightin}. The SVM classifier gives the score as toxic (1) or non-toxic (0). 
%To validate the model, they use $f_{0.5}$ score. 

\subsubsection{Deep  Pyramid Convolutional Neural Networks (DPCNN)}
Johnson and Zhang proposed a deep neural network based model for toxic text classification, which they named as Deep  Pyramid Convolutional Neural Networks (DPCNN) \cite{johnson2017deep}. DPCNN outperformed prior state-of-the-art models on six benchmark datasets  in both sentiment and toxicity classification. 
%The model has 15 weight layers. They proposed a method where the computation is halved after each convolution layer and the model is shaped as a pyramid. For text classification, they ensemble their model with an unsupervised embeddings (two-views embeddings) \cite{johnson2015semi}.
A DPCNN implementation also performed well in the 2018 Kaggle classification challenge~\cite{kaggle2018} by achieving an AUC score of 0.98.  
Similar to the PPA, DPCNN also provides the probability of a text being toxic from 0 to 1. In our evaluation, we use a DPCNN model trained using the Jigsaw dataset. Similar to the PPA scores, we used the threshold of 0.5 to  consider the DPCNN classification as either toxic or non-toxic. 

%\footnote{\url{https://www.kaggle.com/c/jigsaw-toxic-comment-classification-challenge/notebooks}} in Kaggle competition. The training dataset in Kaggle contains  159,572 comments from online. Raters classify each comment into  six  categories  (toxic,  severe\_toxic,  obscene,  threat,  insult,  identity\_hate). The  category  score  for  a  comment  is 1 if the comment belongs to that particular category; the score is 0 otherwise. We use FastText word embeddings to train and test the data in our model \cite{grave2018learning}. 
%We train the DPCNN model with labeled trained dataset and it gives the accuracy by AUC score 0.980967. After training with Jigsaw dataset, we test the model with our two software engineering labeled datasets. For each comment, the model generates the scores for six categories. We only consider the score of toxic category. If the toxic score is more than or equal 0.5, we consider that the comment is classified as toxic; otherwise non-toxic. 

\subsubsection{BERT with fast.ai (BFS)}
Devlin et al. proposed a  pretrained deep bidirectional representations from
unlabeled text language model called Bidirectional Encoder Representations from
Transformers (BERT) \cite{devlin2018bert}.  BERT based models are currently considered as the  state-of-arts in natural language processing (NLP) classification tasks . 
Kurita et al. proposed a toxicity classification model~\cite{kurita2019towards} by fine tuning a  BERT model for  fast.ai library\footnote{\url{https://www.fast.ai/}} for the \footnote{\url{https://www.kaggle.com/keitakurita/bert-with-fastai-example}}. %BERT is pretrained as a deep bidirectional Transformer.
  A BFS model trained using the Jigsaw dataset achieved one of the highest public AUC scores  in the Kaggle competition~\cite{kaggle2018}. 

In our evaluation, we use the publicly available python implementation of the BFS kernel \footnote{\url{https://www.kaggle.com/keitakurita/bert-with-fastai-example/}}. We  retrain a BFS model with the Jigsaw dataset and were able to achieve the  an AUC score of 0.9853 (i.e., same as the listed public score). As BFS also outputs the probability of a text being toxic from 0 to 1, we use the threshold of 0.5 to classify each text as either `toxic' or `non-toxic' based on its BFS score.

%We also follow the similar procedure for training. As fastai is built in pytorch, we train the model in pytorch environment. In the preprocessing step, we follow to use wordpiece tokenizer from BERT. We also wrap the internal tokenizer of fastai on its tokenizar class. We use the vocabulary of BERT toenizar using fastai object. We put it all together in databunch. Using fastai techniques, we have initialized the bert model with loss fuction. Then we pass them to learner and fastai has its' learning rate finder.  As BERT paper mentioned learning rate 3e-5, we have used the same learning rate. Then we have started to train the actual model. It takes around 12 hours to train the model with Jigsaw dataset. After training, we get get a public score of 0.98553 and a private score of 0.9858 \url{http://mlexplained.com/2019/05/13/a-tutorial-to-fine-tuning-bert-with-fast-ai/}.  

%After the training, we test the model with our two datasets of SE. We follow the simial approach like DPCNN to find the accuracy of BERT with Fastai model. 

\subsubsection{Hate Speech Detection (HSD)}
Davidson et al. proposed an automated multi-class classifier to classify a text as either a hate speech, or an offensive language or neither~\cite{davidson2017automated}. The HSD model was classified using a dataset of 25K manually labeled tweets and achieved a precision 0.91, recall of 0.90, and F1 score of 0.90 during five-fold cross validations. 
During our evaluation, we use the pretrained HSD models and classify a text as `toxic' if it was classified as either `a hate speech' or `an offensive language' by the HSD model.

\subsection{Evaluation Metrics}
%\anote{Talk about the metrics that we would be using to compare the tools }. 
In our evaluation, we use the following five measures to compare the performances of the tools on our SE datasets.

\begin{itemize}
\item \textbf{Accuracy:} Accuracy is the ratio of texts that were correctly classified by a tool.
\item \textbf{Precision:} The ratio  between the number of toxic texts that are correctly classified by  a tool and the number of of texts that are marked as toxic by the same tool.
\item \textbf{Recall:} The ratio between the number of toxic texts that are correctly classified by a tool and the number of toxic texts in the dataset.
\item \textbf{F-score:} The harmonic mean of precision and recall.
\item \textbf{Cohen's Kappa ($\kappa$):} The level of agreement between a tool's classification and our manual labeling as measured using Cohen's Kappa ($\kappa$)~\cite{cohen1960coefficient}.
\end{itemize}

We also calculate the level of agreement between each pair of tools using the Cohen's Kappa ($\kappa$) \cite{cohen1968weighted} to determine how obtained results would vary if a different tool was selected for analyses.

%We have evaluated the six models after testing by our SE dataset. We consider Precision, Recall, F score, and Accuracy as evaluation metric for six tools.  We chose those metrics because some sentiment analysis evaluation tools also use the same metrics \cite{barbieri2016overview} \cite{novielli2018benchmark}. Precision is the measurement where we calculate the ration of true positive and all classified comments as positive. On the other hand, recall represents the ratio of true positives and actual positives. F-score is the harmonic mean of precision and recall. 

\section{Results}
\label{sec:evaluation}
The following subsections present the results of the three research questions introduced in the Section~\ref{sec:intro}.

\subsection{RQ1: How do Contemporary Toxicity Detectors Perform on an SE Dataset?}
%\anote{Put the results as a table. There should be two tables. i) provides precision/recall and f-score of each tool. ii) provides the agreements between a pair of tools. }
%\begin{table*}
%	\caption {Performance of  the five Toxicity Analysis Tools on our Kaggle Test dataset} 
%	\centering \label{tables:testing_accuracy}
%	\input{Tables/testing_dataset_accuracy}
%	%\vspace{-5pt}
%\end{table*}

\begin{table*}
	\caption {Performance of  the five Toxicity Analysis Tools on the three datasets} 
	\centering \label{tables:testing_accuracy}
	%\resizebox{\linewidth}{!}{

\begin{tabular}{|l|l|r|r|r|r|r|}
\hline
\textbf{Dataset} & \textbf{Tools}                       & \textbf{Precision ($p$)}      & \textbf{Recall($r$)}         & \textbf{F-Score ($f$)} & \textbf{Accuracy ($A$)}       &  Kappa (\textbf{$\kappa$)}              \\ \hline

\multirow{5}{3.5cm}{ Jigsaw Sample \textit{(baseline)} }     & PPA & 0.762 & 0.986 & 0.858 & 0.893   & 0.775 \           \\ \hhline{~------}

        & STRUDEL & 0.734          & \textbf{0.990} & 0.843    & 0.877         & 0.746           \\ \hhline{~------}
        & DPCNN  &\textbf{0.896}         & 0.829          & \textbf{0.861}   & \textbf{0.911} & \textbf{0.796}  \\ \hhline{~------}
        & BFS            &  0.887         &  0.834          &  0.859   &   0.909         & 0.793             \\ \hhline{~------} 
        & HSD       & 0.889 & 0.427          & 0.577   & 0.791         &  0.461         \\ 
          \hline

\multirow{5}{*}{Code Review}     & PPA & 0.397 & 0.762 & \textbf{0.522} & 0.720          & \textbf{0.351}          \\ \hhline{~------}
        & STRUDEL & 0.347          & \textbf{0.861} & 0.495   & 0.648          &  0.294          \\ \hhline{~------}
        & DPCNN  &\textbf{0.708}         & 0.285          & 0.406   & \textbf{0.833} & \textbf{0.33} \\ \hhline{~------}
        & BFS            & 0.663          & 0.253          & 0.366   & 0.824          &   0.287          \\ \hhline{~------} 
        & HSD       & 0.705 & 0.051          & 0.095   & 0.805          & 0.071           \\ 
          \hline
       
\multirow{5}{*}{Gitter Ethereum} & PPA & 0.707 & 0.806 & \textbf{0.753} & \textbf{0.813} & \textbf{ 0.604 } \\ \hhline{~------} 
        & STRUDEL & 0.626          & \textbf{0.880} & 0.732   & 0.771          & 0.542          \\ \hhline{~------} 
        & DPCNN                       & 0.901          & 0.511          & 0.652   & 0.806          &  0.532         \\ \hhline{~------}
        & BFS            & 0.892 & 0.524          & 0.660   & 0.809         & 0.539          \\ \hhline{~------}
        & HSD   & \textbf{0.978}         & 0.238          & 0.382   & 0.728          &  0.283          \\ \hhline{~------}
        \hline
\end{tabular}
%}
	%\vspace{-5pt}
\end{table*}

%\begin{figure*}
%	\centering  \includegraphics[width=\linewidth]{Figures/tool-performance.eps}
%	\caption{Performance of the six toxicity detectors}
%	\label{fig:perforamce}	
	
%\end{figure*}

Table ~\ref{tables:testing_accuracy} shows the precision, recall, F-score, and accuracy of the five tools, when evaluated on our three datasets. 
On the baseline Jigsaw sample test dataset, DPCNN achieves the best precision, F-score, and accuracy, while STRUDEL achieved the best recall. Four out of the five tools (i.e., except HSD) achieved `substantial' agreement with the human raters. 

On the code review dataset, the STRUDEL tool, which is customized for the SE domain, achieves the best recall. However, due to lower precision than the PPA, STRUDEL falls behind in terms of both F-score and accuracy. Although, HSD had the the second highest precision among the five tools on the code review dataset, it also had the lowest F-score due to its failure to identify (i.e., false negatives) toxic texts  that do not express hate speeches. DPCNN had the highest precision, accuracy and kappa on the code review dataset. The kappa values suggest that the five tools achieved at best  `Fair'  agreements with the human raters and therefore may not suitable to identify toxic texts from code review interactions.

All the five tools performed better on the Gitter dataset than each performed on the code review dataset, since the Gitter dataset has higher ratios of toxic comments as well as higher number of messages with profanities. Similar to the code review dataset, the STRUDEL achieved the best recall on the Gitter dataset but falls behind the PPA in terms of both Accuracy and F-Score. Both the BFS and the HSD achieved high precisions, but failed to achieve high F-scores due to large number of false negatives. The kappa values suggest that four out of the five tools achieved `moderate' agreements with the human raters, which can be considered as improvements over the performances achieved by those tools for the code review dataset. 

%With an F-score of 0.79, PPA can be somewhat useful in identifying toxic texts from developer chats, although, this value is significantly below the performance achieved by the state-of-the-art toxicity detectors on the Jigsaw or OffensiveEval datasets~\cite{kurita2019towards,zampieri2019semeval}.
By comparing each tools performance on the two SE datasets against its performance on the Jigsaw sample, we noticed significant  degradations of F-scores. Both precisions and recalls of each tool dropped by more than 0.10 on the two SE datasets.
Among the five tools, PPA provides the best F-scores on both of our SE datasets and may be considered as the baseline for building SE domain specific toxicity detectors. 

%Gitter Ethereum dataset contains the developers communication messages those are likely as public communication. Developers communication contains many comments those are used in online conversation. BERT with Fastai gives the highest Precision (0.9886) for testing the Gitter Ethereum dataset. Like code review dataset, STRUDEL\_toxicity\_detector has achieved the highest recall (0.9477) for Gitter Ethereum. Similarly, Perspective API shows the highest F1-score (0.7903) for developers communication dataset. Moreover, Perspective API achieves the highest accuracy (0.7789) among all the tools for Gitter Ethereum dataset.

%We have found that Hate Speech Detection tool outperforms other five tools for it`s precision (0.8433). This tool detects a small number of false positives. This tells us that it rarely detects the non-toxic comments as toxic. STRUDEL\_toxicity\_detector has the highest recall (0.9113) among six tools for testing the Code Review dataset. STRUDEL\_toxicity\_detector tool is designed and trained for SE datasets. For that reason, it rarely detects toxic comments as non-toxic. But this tool has a low precision value (0.3827). Goodle`s Perspective API gives the highest F-score (0.5403) among all tools. Perspective API is pretrained by Google Developers with a large number of public domain comments. DPCNN is trained by Jigsaw 2018 toxic comments train dataset. 

Table \ref{tables:agreement} shows the agreements between each pair of tools measured using the Cohen's Kappa ($\kappa$). Since we have five different tools, there are ten possible pairings. The highest level of agreements were seen between the PPA and STRUDEL pairs on the both SE datasets. Based on the $\kappa$ values, agreements between these two tools can be considered as `substantial'. Since STRUDEL uses PPA scores as an input for classification,  `substantial' agreements between these  two tools are not surprising.  The only other pair that showed `substantial' agreement is the DPCNN-BFS pair on the Gitter dataset, since both DPCNN and BFS models are trained using the same dataset (i.e., Jigsaw toxicity dataset), their agreements on the Gitter dataset may not  be surprising, while their `moderate' level of agreement on the code review dataset deserves more investigation.  The lowest level of agreement was obseved between the STRUDEL-HSD pair.
%The HSD and BOM had lowest level of agreements with the other tools and in some cases negative Kappa values suggest significant level of disagreements from BOM with the other tools.  Among the the fifteen possible pairings, only one pair (i.e., STRUDEL -PPA ) shows substantial agreements on both datasets. 
These results suggest that if an empirical investigation of toxic texts are conducted using one of these tools, the results may be different if we select one of the different tools, except for switching between STRUDEL and PPA may still yield the same results.

%Although DPCNN achieves high accuracies and F-scores on both datasets, the level of agreements between DPCNN - PPA / DPCNN -STRUDEL pairs  is `Fair'. 
%Table \ref{tables:agreement} (a) shows the agreement with our labeling with six tools. We have found that DPCNN has the highest agreement (76.1\%) with our manual labeling for code review dataset. BERT with Fastai achieves the highest agreement (76.9\%) for Gitter Ethereum dataset.

\begin{table}
	\caption {Level of agreements between the tool pairs on our SE datasets} 
	\centering \label{tables:agreement}
	\resizebox{\linewidth}{!}{
% Please add the following required packages to your document preamble:
% \usepackage{multirow}

\begin{tabular}{|p{2.5cm}|r|r|r|r|r|r|}
\hline
\multirow{2}{*}{\textbf{Dataset}}        &  \multicolumn{6}{c|}{\textbf{{Level of agreement between tool pairs ($\kappa$)}}}                    \\ \hhline{~------}
                               
                                 & \multicolumn{2}{l|}{}        & \textbf{STRUDEL}    & \textbf{DPCNN}   & \textbf{BFS}     & \textbf{HSD}          \\ \hline
\multirow{6}{2.5cm}{\textbf{Code Review}}     & \multicolumn{2}{l|}{\textbf{PPA}}     & \textbf{0.777}       & 0.225   & 0.221    & 0.033       \\ \cline{2-7} 
                                 & \multicolumn{2}{l|}{\textbf{STRUDEL}} &            & 0.156   & 0.151   & 0.023      \\ \cline{2-7} 
                                 & \multicolumn{2}{l|}{\textbf{DPCNN}}   &            &         & 0.595    & 0.18       \\ \cline{2-7} 
                                 & \multicolumn{2}{l|}{\textbf{BFS}}     &            &         &         & 0.142      \\  \hline

\multirow{6}{2.5cm}{\textbf{Gitter Ethereum}} & \multicolumn{2}{l|}{\textbf{PPA}}     &\textbf{0.812}      & 0.497   & 0.504   & 0.236       \\ \cline{2-7} 
                                 & \multicolumn{2}{l|}{\textbf{STRUDEL}} &            & 0.383    & 0.39     & 0.168      \\ \cline{2-7} 
                                 & \multicolumn{2}{l|}{\textbf{DPCNN}}   &            &         & 0.759  & 0.49       \\ \cline{2-7} 
                                 & \multicolumn{2}{l|}{\textbf{BFS}}     &            &         &         & 0.481       \\  \hline
\end{tabular}

}
	%\vspace{-5pt}
\end{table}

%We show the pair of agreement of each tool in table \ref{tables:agreement} (b). We have found that Perspective API agrees more than 90\% with STRUDEL\_toxicity\_detector . It seems similar because STRUDEL\_toxicity\_detector is built with the score of Perspective API. On the other hand, DPCNN and BERT with Fastai agrees 87.1\% for scoring the code review dataset and 90.3\% for Gitter Ethereum. We have trained both DPCNN and BERT with Fastai using Jigsaw dataset. For the same training dataset, the score would be similar. On other cases, tools shows a large number of disagreements. It gives us a suggestion that we need a customized tool for SE dataset. 

\begin{boxedtext}
\textbf{Finding 1:} \emph{While contemporary toxicity detectors have moderate agreements with human raters on identifying toxic texts from informal conversations such as chat messages, they perform poorly on a more formal SE conversation such as code reviews. Since only one out of the ten possible pairs had substantial agreements with each other, the results of an empirical study may significantly differ, if we switch from one tool to another from those nine low agreement pairs.}
\end{boxedtext}

\subsection{RQ2: What are the Categories of SE Texts that Contemporary toxicity Detectors are More Likely to Misclassify?}
\label{sec:rq2}
We conducted secondary investigations to identify cases where most of  the tools misclassified to identify the challenges in developing an SE domain specific toxicity detector.

%We have looked the insight of the results of each tool for finding why they can not give a good score for toxic content classification in our SE dataset. 
%We are trying to identify the challenges for the tools to classify the toxic comments in SE dataset. We manually investigate the score for each comments where each tools can not give the correct score with our manual labeling. Toxicity detection from a context is a part of NLP where it varies from person to person. 
All the tools used in our study are based on supervised models, pretrained with a large labeled datasets. However, many of the words has different meanings in the SE context than in general English. Most of the tools failed for such words. In the following we list such words with examples.

\begin{itemize}
    \item \textbf{kill:}  is frequently used during code reviews and developer chats to suggest killing a process or simply removing a code snippet. For example,  ``\textit{yeah, they don't seem to be needed, so let's kill them.}'', suggests removing some unnecessary code snippet, which was misclassified as toxic by most of the tools.
   
   % Though STRUDEL tool worked on removing non-toxic SE phrases, they also miss-classify the text. Sometimes those phrases are also used as toxic comment in SE domain. For that reason, it is important to find out the meaning of full text to classify it as toxic or non-toxic. 
    \item \textbf{execute:}  refers to running a process or application in the SE domain. For example,`` \textit{Any program executed by any kernel thread, including usermodehelper, from rootfs will switch to init?}'', refers to running a program, but was incorrectly classified as toxic.

    \item \textbf{die, dead:} Both `die' and `dead' refers to state of program execution or code snippets and are often misclassified by off-the-shelf toxicity detectors. For example, \textit{``Remove the old, dead code.}''
    
    \item \textbf{garbage:}  is another word that can be used both in toxic and non-toxic ways. For example, in ``\textit{initialize init\_pid\_ to -1 here so it doesn't have garbage in it}'', garbage cannot be classified as toxic. However, a developer referring another developer's code as `garbage' would be toxic. We noticed `garbage' used in non-toxic contexts for most of the cases and were misclassified by the tools.

    \item \textbf{dummy:} is often used to refer to placeholder files or objects.  ``\textit{What is it used for? An empty dummy file should work.}'', is an example of misclassification of a text with this word.
    
    \item \textbf{junk:} Under slang terms, `junk' refers to male privates. However, in the SE domain `junk' often refers to useless objects and can be misclassified. For example, ``\textit{I'd like to have that here too, since input may have junk data after a valid CBOR.}''
    
    \item \textbf{dirty:} In the SE domain, `dirty' often refers to a modified file or memory location. A misclassification with the word dirty:,  ``\textit{why not place this in the dirty bits iteration? (with a comment on why we need it for D3D11)}''.
    
    \item \textbf{trash:} refers to removing file, code snippet or objects. A misclassified example is, ``\textit{You really don't need to derive from std::less<>. If anything, you should be deriving from std::binary\_function, but it is really not needed for std::set to work correctly, so I would just trash that base class.}''
    
    \item \textbf{daemon:}  refers to  a computer program that runs as a background process in the SE domain. However, in a non-SE domain, it may refer to something supernatural and therefore, was classified as toxic. An example  of such occurrence is : ``\textit{Based on the old version, it looks like lxc should be built even if USE=daemon is not sent. ...}''
    
    \item \textbf{naked:} is usually considered as a toxic word with a sexually explicit reference. However, in the C programming language a `naked pointer' refers to pointers that  can be used to point to another object. Texts with the word `naked' was frequently misclassified during code reviews. For example, \textit{``For now, let's keep it like this, there's a discussion going on what to do with the naked C++ pointers.}''
    
    \item \textbf{dump:} can be used as a slang to indicate `the act of defecation'. However, in the SE context dump often refers to storing data.  For example, an example of misclassification with dump is: ``\textit{Use json.dump, json.load instead of doing your own string parsing. ...}''
    
    \item\textbf{stupid, dumb, idiot, fool, ignorant:} Most of the classifiers marked all the texts with these words as toxic. However, during both code reviews as well as Gitter chats, developers frequently used those words to express humility. For example, ``\textit{Maybe a stupid question: where's this variable defined?}''.
    
    \item \textbf{CAPITALIZED ACRONYMS:} In C or C++ constant variables are often declared in all caps. References to code segments are often included in code reviews.  For example, ``\textit{Make this another DCHECK.}'' Some of the tools incorrectly marked unknown capitalized acronyms as toxic.
    
\end{itemize}

\begin{boxedtext}
\textbf{Finding 2:} \emph{Many of the words, that are used under toxic contents in non-SE domains, have different meanings in the SE context, are more frequently misclassified by the toxicity detectors.  }
\end{boxedtext}

\subsection{RQ3: Does Retraining on a SE Dataset Improve the Performances of Contemporary Toxicity Detectors?}
%\anote{How do tools perform if we retrain those on our dataset? We cannot retrain all tools. Which tools are retrainable? Does retraining changes performance?}
While we intended to to reevaluate all the five tools after retraining those on a SE dataset, it was feasible for us to retrain only two models (i.e., DPCNN and BFS) with our dataset. We could not retrain the PPA, since its source code is proprietary. STRUDEL uses the PPA model and the dataset to customize the Stanford politeness detector for STRUDEL is not publicly available. HSD is a multiclass model with three classes (`hate speech', `offensive', and `neither'). Since our datasets are not labeled accordingly, we excluded the HSD model. 
%Although, we were able to use the pretrained BOM models, we were unable to retrain it from scratch using our dataset.

%Both DPCNN and BFS use neural network based models. Since deep neural network models perform better if trained on  a large dataset, we create a single dataset 7,162 labeled texts by combining the two datasets. 

We evaluated the the models using 10-fold cross-validations. 
Table~\ref{tab:retraining} shows the average performances of the two models after retraining on our datasets. 
Both of the models achieves significant performance improvements after retraining on our SE dataset. The DPCNN based model achieved an F-Score of 0.88 on the code review dataset, which is better than its baseline performance (i.e., F-Score of 0.86 on the Jigsaw sample). But it under-performed on the Gitter dataset with an F-Score of 0.731. On the other hand, the BFS based model's F-Score of 0.860 on the Gitter dataset was almost similar as its baseline  performance (F-Score of 0.859). However, BFS under-performed on the code review dataset with an F-Score of 0.731.
%We evaluated the DPCNN with 10 fold cross validation of our SE dataset.
%We have found that DPCNN improves precision, recall, f1 score and accuracy while evaluating with SE dataset. 

%We have evaluated the BFS model by splitting the combined SE dataset where 80 percent is used for training and 20 percent used for testing the model. 

\begin{boxedtext}
\textbf{Finding 3:} \emph{Both models achieved significant performance boosts after retraining on our SE datasets. Two out of the four models beat its' baseline F-Scores achieved on a non-SE dataset. A large scale labeled toxicity dataset of SE interactions may enable developing SE domain specific toxicity detectors that can be used for identifying toxic texts from real-world SE interactions. }
\end{boxedtext}

\begin{table*}

\caption {Performance of the toxicity detectors after retraining on our SE dataset} \label{tab:retraining} 
\centering
%\resizebox{\linewidth}{!}{
\begin{tabular}{|l|l|r|r|r|r|r|}
\hline
\textbf{Dataset} & \textbf{Tools}                       & \textbf{Precision ($p$)}      & \textbf{Recall($r$)}         & \textbf{F-Score ($f$)} & \textbf{Accuracy ($A$)}       &  Kappa (\textbf{$\kappa$)}              \\ 
\hline \hline
\multirow{2}{*}{Code Review}     & DPCNN & \textbf{0.880} & \textbf{0.890} & \textbf{0.880} & \textbf{0.920}         &   \textbf{0.817}        \\ \hhline{~------}
        & BFS &  0.780          &  0.688  &  0.731   & 0.898           &  0.669           \\ %\hhline{~------}
          \hline
          \hline
       
\multirow{2}{*}{Gitter Ethereum} & DPCNN & \textbf{0.840}   & 0.670   & 0.740 & \textbf{0.910}  & 0.692 \\ \hhline{~------} 
        & BFS & 0.838  & \textbf{0.884}  & \textbf{0.860}    &  0.892          & \textbf{0.773}           \\ %\hhline{~------} 
        \hline
\end{tabular}
%}

\end{table*}

\section{Implications} \label{discussion}
%\anote{What do we learn from our evaluation? What are the guidelines for researchers? What are the instructions for tool developers? }

In this paper, we evaluated five contemporary toxicity detectors on two SE datasets.  Following are the key lessons obtained from this study.
\begin{enumerate}
\itemsep5pt    

\item \textbf{  Off-the-shelf tools are reliable in identifying profanities.} Profanities are the most common sources of  online toxicities. We found most of the tools highly reliable in flagging texts with profanities. Therefore, if an SE community only wants to flag profane languages, off-the-shelf tools such as PPA can be useful with its profanity detection model.

\item \textbf{  Off-the-shelf tools are not reliable on SE datasets.}
Although, PPA and STRUDEL show moderate performance on  identifying toxic texts from Gitter messages, their performances are not reliable on  formal conversations such as code reviews. Therefore, off-the-shelf tools must be evaluated for reliability on a dataset drawn from the study context before their application.

\item \textbf{  Retraining off-the-shelf tools on a SE dataset significantly improves performance.}
While we conducted a preliminary investigation by retraining two of the off-the-shelf tools on our SE dataset, the results are highly promising. We believe, if we retrain contemporary models using a larger and more robust SE dataset than the one used in this study and add SE domain specific preprocessing, we can develop a reliable toxicity classifier for the SE domain.

\item \textbf{  SE domain specific preprocessing may improve performances.}
We noticed that several misclassifications from the existing tools were due to code snippets included in developer communications. Since SE domain specific sentiment analysis tools also recommend filtering out code snippets~\cite{ahmed2017senticr}, we believe  that preprocessing steps to identify and remove code snippets may improve the performances of SE domain specific toxicity detectors.

\item \textbf{  Excluding SE domain specific words may cause false negatives.}
The results of RQ2(Section~\ref{sec:rq2}) illustrated several words that may have different meaning in an SE context. While using the approach adopted by Raman et al.~\cite{ramanstress}, we can replace these words with a more neutral words, and reduce those misclassifications, this approach may also generate false negatives if these words are truly used to express a toxic opinion. For example, following lists shows toxic usage of those words from our dataset:
\begin{itemize}
    \item garbage: ``\textit{Why you changed this to \%ecx? it is garbage here.}''
    \item kill: ``\textit{go kill yourself}''
    \item junk: ``\textit{so I don't have to clean up my junk after myself }''
    \item dirty: ``\textit{Only if you promise to talk dirty to me}''
    \item dump: ``\textit{I wouldn't recommend telling girls they are pretty, I mean if it's the first or second thing you say your intentions are clear, you just want sex, and then they dump you.}''
    \item die: ``\textit{well the US can go and die}''
    \item dead: ``\textit{... should just move over to ethereum immediately... no point in flogging a dead horse}''
    
\end{itemize}

\item \textbf{  A reliable toxicity detector must identify the target of words to identify expressions of humility.}
Sentences using the words: `idiot', `stupid', `dumb', `ignorant', and `fool' to express humility were often misclassified by the contemporary toxicity detectors. We found both toxic and non-toxic usages of those words. Since during expressions of humility, these words refer to the author him/her self or his/her works, a reliable toxicity classifier must identify the target of those words to identify toxic contexts from non-toxic ones.

  \end{enumerate}

\section{Threats to Validity} \label{threats}

The first threat to validity for this study is our selection of data sources which come from four FOSS projects. While these projects represent four different domains, many domains are not represented in our dataset. Moreover, our projects represent some of the top OSS projects with organized governance. Therefore, several categories of highly offensive texts are underrepresented in our datasets. 

Second, our stratified sampling strategy was based on the scores provided by the PPA. Although, we manually verified all the texts classified as `toxic' by the PPA, we randomly selected only 5,510 texts that had PPA scores of less than 0.5. Among those texts, we identified 513 toxic texts (9.3\%). Therefore, if the PPA misclassified some categories of `toxic' comments and also our random selections missed those, instances of such texts may be missing in our datasets.

%Third, since we have empirically developed our rules to manually label toxic contents, we may be missing some rules, where representative conversations illustrating needs for these rules were absent in our datasets.

Third, we accepted the default parameters for the selected tools and did not use parameter tuning to improve performances. Therefore, some of the tools may have achieved better performances on our datasets through parameter tuning.

Finally, although we have selected a diverse set of tools trained on different datasets, we may be missing some tools that could have achieved a better performance on our datasets. To enable evaluations of more tools on our datasets, we had made those publicly available on a Github repository.

%\anote{What are threats to validity?}
\section{Conclusion} \label{conclusion}
In this paper, we empirically evaluated STRUDEL, the only  SE domain specific toxicity detector, as well as four other state-of-the-art general purpose toxicity detectors on two labeled SE datasets. We empirically developed a rubric to manually label toxic SE interactions, and using this rubric, we manually labeled a dataset of 6,533 code review comments and 4,120 Gitter messages. 

 The results of our analyses suggest that none of the contemporary toxicity detectors could achieve adequate performances to justify practical applications. The performances of the tools included in our study dropped more significantly on a formal SE communication dataset such as code review than on a dataset of informal communication such as Gitter messages. The significant disagreements between most of the tool pairs also suggest that results of an empirical study using one of these tools may differ significantly if we switch from one tool to another one. 
 One of the primary limitations of existing tools are their failures to identify non-toxic contexts of certain words that are commonly used in toxic contexts in a non SE-domain but may have different meanings in the SE domain. Sentences with source code snippets  and with the words, such as: `idiot', `stupid', `dumb', `ignorant', and `fool' to express humility were also frequently  misclassified.
  We retrained two of the models from our study on our SE dataset and obtained  highly promising results with two out of the four models beating its' baseline F-Scores obtained on a non-SE dataset. These results suggest that the development of a highly reliable SE domain specific toxicity detector is feasible by retraining existing models on a large-scale and robust labeled dataset of SE interactions.

Based on our investigations, we have identified several key lessons that may help researchers in developing an SE  domain specific toxicity detector. The rubrics developed in this study to manually label toxic SE interactions as well as the two labeled datasets, which are publicly available, will be also helpful. The future direction for this research include: i) the development of a large-scale and robust labeled dataset, ii) the development and evaluation of SE domain specific text preprocessing steps to improve the performances of toxicity classifiers, and iii) the development of a reliable toxicity classifier for the SE domain.
%We have plotted a study on the performance of contemporary toxic content classifiers in software engineering dataset. We have used our manually labeled two SE dataset those can be used for further research. During the evaluation, we compared the six tools performance and we manually investigated the failure of each tools in SE dataset. We mentioned the failures in error analysis section. Moreover, we reported the further direction for researcher to build SE specific toxic content classifiers in discussion setion. We have found that reliable classifier can be achievable if it is trained by the SE specific large dataset. We have contributed a rubric for detecting a toxic content in SE domain. Those rules would be helpful to build a toxic content classifier in SE community. Our study suggest the people in software engineering domain to make healthy interactions with his/her peers. For further research, we will work on larger dataset than we manually labeled and we plan to contribute our study to develop a toxic content classifier in SE domain.  

%%
%% The next two lines define the bibliography style to be used, and
%% the bibliography file.
\bibliographystyle{IEEEtran}
\bibliography{toxicity-references}

\end{document}